\documentclass[twocolumn,showpacs,floatfix]{revtex4}

\usepackage{graphicx}

\begin{document}

\newlength{\plotwidth}          
\setlength{\plotwidth}{7.5cm}


\title{\large Combined atomic force microscope and electron-beam lithography \\ used for the fabrication of variable-coupling quantum dots}

\author{M. C. Rogge}
\email{rogge@nano.uni-hannover.de}
\author{C. F\"uhner}
\author{U.~F. Keyser}
\author{R.~J. Haug}
\affiliation{Institut f\"ur Festk\"orperphysik, Universit\"at
Hannover, 30167 Hannover, Germany}
\author{M. Bichler}
\author{G. Abstreiter}
\affiliation{Walter Schottky Institut, TU M\"unchen, 85748
Garching, Germany}
\author{W. Wegscheider}
\affiliation{Angewandte und Experimentelle Physik, Universit\"at
Regensburg, 93040 Regensburg, Germany}

\date{\today}

\begin{abstract}
We have combined direct nanofabrication by local anodic oxidation
with conventional electron-beam lithography to produce a parallel
double quantum dot based on a GaAs/AlGaAs heterostructure. The
combination of both nanolithography methods allows to fabricate
robust in-plane gates and Cr/Au top gate electrodes on the same
device for optimal controllability. This is illustrated by the
tunability of the interdot coupling in our device. We describe our
fabrication and alignment scheme in detail and demonstrate the
tunability in low-temperature transport measurements.
\end{abstract}

\maketitle


Quantum dots realized in various semiconductor materials have
received great interest for the last decade~\cite{Kouwenhoven-97}.
They are often called artificial atoms for the similarity of their
zero-dimensional electronic spectra with their real
counterparts~\cite{Kastner-92}. In an expansion of this concept
two or more coupled quantum dots in close spatial vicinity form
artificial molecules~\cite{Wiel-03}. Quantum dot molecules allow
to precisely control properties like electron number, quantum
mechanical state or interdot coupling by experimental
parameters~\cite{Wiel-03,Kouwenhoven-01a,Blick-98a,Holleitner-02}.
This tunability makes them promising candidates for the
realization of quantum computers based on electron spin (see
e.~g.\ Refs.~\cite{Loss-98,Hu-00,Golovach-02}). Recently, concrete
steps towards a practical realization of this concept have been
proposed~\cite{Vandersypen-02}.


\begin{figure}
 \includegraphics{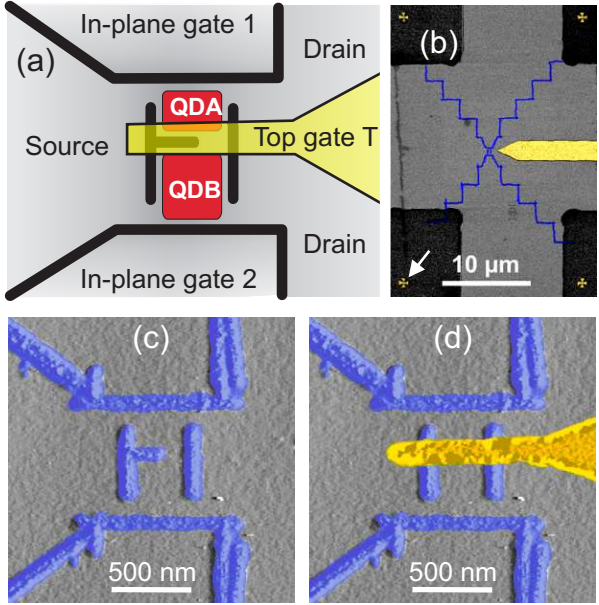}
 \caption{(a) Schematic picture of our sample with two quantum
dots QDA and QDB. The top-gate controls the coupling of the dots
without depleting the 2DES in the drain area. (b) Colorized SEM
picture after the local oxidation step (2) including mesa (grey),
inner marker set (4 yellow crosses, the one in the lower left
corner is highlighted by a white arrow), coarse part of the gate
(yellow), and oxide lines (blue). The outer marker set is not
shown. (c) Magnified colorized AFM image of the double-dot
structure. (d) The same sample after adding the top-gate (yellow)
by e-beam lithography.}
 \label{fig1}
\end{figure}

We focus on the fabrication of coupled quantum dots with both dots
connected to common leads. With conventional electron-beam
(e-beam) lithography such devices were realized either with only
one dot connected to the leads~\cite{Hofmann-95} or with both dots
connected to separate contacts~\cite{Molenkamp-95,Adourian-99}.
Only recently Holleitner {\em et al.} succeeded in the fabrication
of parallel double quantum dots connected to common source and
drain contacts by e-beam
lithography~\cite{Holleitner-01,Holleitner-02}. They used an
additional e-beam patterned calixarene spacer
layer~\cite{Fujita-96} to locally decrease the depletion created
by top-gate electrodes in a two-dimensional electron system
(2DES).

In our fabrication scheme, we combine conventional e-beam
lithography and local anodic oxidation (LAO) of a 2DES using an
atomic force microscope (AFM)~\cite{Ishii-95,Held-98,Keyser-00}.
LAO allows to pattern static insulating lines into the 2DES. These
lines form the basic structure including in-plane gates and
tunneling barriers.  LAO based structures are easier to produce
and operate while being less sensitive to electrostatic discharges
compared to generally more complex top-gate structures involving
many gates. In addition, the potential walls created by LAO are
extremely steep~\cite{Fuhrer-01b}.  We add few metallic top-gate
electrodes by e-beam lithography to improve the tunability of the
basic structure. In our double quantum dot sample, a LAO based
tunneling barrier situated below a top-gate electrode allows to
locally increase the depletion created by the gates. Thus our
combination of LAO and e-beam lithography allows to enhance the
tunability of LAO based structures without introducing the
disadvantages and added complexity of equivalent purely top-gate
based structures.


In this letter we present our nano-fabrication scheme in detail.
We demonstrate its feasibility by creation of a parallel quantum
dot molecule with tunable interdot coupling.  Finally, we employ
low-temperature transport measurements on this device.


An outline of our structure is shown in Fig.~\ref{fig1}(a). Two
LAO based quantum dots are connected via 80--90~nm wide point
contacts to the source and drain reservoirs. In-plane gates 1 and
2 allow to tune the electrochemical potential of the adjacent dot
and tunneling barriers. The dots are connected by a 100~nm wide
opening to allow for tunnel coupling. The opening is covered by a
top-gate electrode on the surface for electrostatic tuning of the
interdot coupling. Due to the added depletion of the tunneling
barrier underneath, the top-gate can be operated at voltages which
are not fully depleting the common 2DES drain contact.


Our fabrication scheme starts with a GaAs/AlGaAs-heterostructure
suitable for LAO.  Our heterostructure contains a two-dimensional
electron system (2DES) with a low temperature mobility of
$\mu=42$~m$^2$/Vs and a sheet density of $n = 5\cdot
10^{15}$~m$^{-2}$ located 34~nm underneath the surface. The layer
sequence consists of (from top to bottom): a 5~nm thick GaAs cap
layer, 8~nm of AlGaAs, the Si-$\delta$-doping, a 20~nm wide AlGaAs
barrier, and 100 nm of GaAs.  Using standard photolithography we
fabricate a mesa and Au/Ge/Ni ohmic contacts. Three
nano-lithographic steps which have to be carefully aligned
relative to each other follow: (1) define a common coordinate
system for the following steps by producing Cr/Au markers using
e-beam lithography, (2) LAO of the basic double-dot with in-plane
gates, and (3) add the metallic finger gate with e-beam
lithography.

In the first step, we apply e-beam lithography to pattern a metal
layer consisting of 7~nm Cr and 30~nm Au for markers. Metallic
alignment markers are needed to define a common e-beam/LAO
coordinate system because the oxide lines produced by LAO are not
visible in the electron microscope when covered by PMMA e-beam
resist.  We use two sets of markers: One set of markers is placed
close to the center of the structure and used to check the
alignment of the later LAO step with the AFM. The other one is
placed farther outside where it is safely scanned with the
electron microscope to align the final e-beam step without
exposing resist in the central dot region.  In addition, we
produce the coarse part of the gate which is used as an additional
marker for LAO. This is shown in Fig.\ \ref{fig1}(b).  The
proximity of the LAO markers to the critical center of the
structure allows to align the later LAO step with a high absolute
accuracy without suffering from distortions of the AFM piezo and a
limited relative AFM resolution at large scanning fields.

For the AFM lithography which is employed in the second
nanolithography step, we use our LAO scheme described in Ref.\
\cite{Keyser-00}. Local oxidation of the heterostructure surface
locally modifies the band structure and depletes the 2DES
underneath.  This directly transfers the oxidized pattern from the
surface into the electronic system.  We fabricate the basic double
dot structure shown in Fig.\ \ref{fig1}(b) and (c). All oxide
lines are about 100~nm wide and create insulating barriers in the
2DES. To align the AFM lithography we use the coarse part of the
gate produced in the previous e-beam step.  After oxidation, we
scan the LAO structure and the inner marker set with the AFM to
check the alignment accuracy and to compensate for a possible
error in the next step.

We complete our sample by adding the 100~nm wide gate finger (7~nm
Cr/30~nm Au) using e-beam lithography.  For alignment we use the
outer marker set from step (1).

The result is shown in Fig.\ \ref{fig1}(d). The figure illustrates
how crucial an exact alignment of e-beam and AFM lithography
relative to each other is.  Since the width of the channel between
the in-plane gates is 800~nm, we need to control the vertical
position of the finger gate with an accuracy which is better than
50~nm. Only an exactly positioned finger gate allows to influence
mainly the interdot coupling instead of the tunneling barriers to
the reservoirs.  This accuracy is achieved with the scheme
described above.


\begin{figure}
 \includegraphics{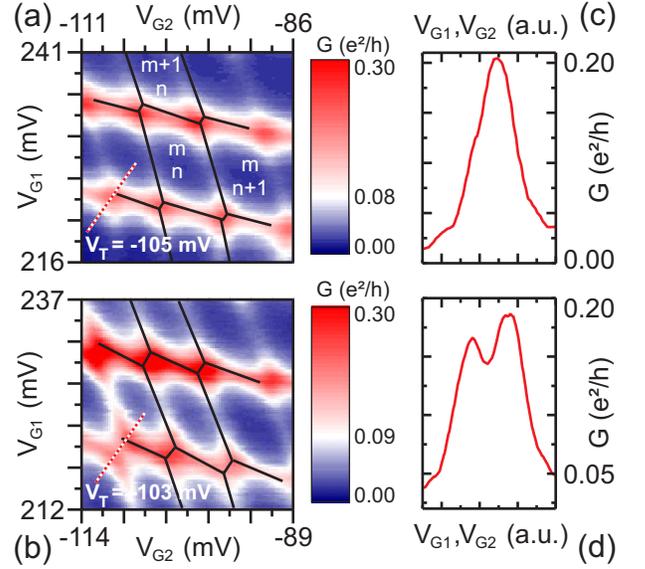}
 \caption{(a), (b) Grey scale plot of the linear conductance
$G$ as a function of in-plane gate voltages $V_{G2}$, $V_{G1}$ for
different interdot coupling strengths. In the hexagonal shaped
Coulomb blockade regions the electron number $(m,n)$ on dots $A$
and $B$ is stable. (c),(d) Traces as marked in (a),(b). The graphs
clearly show the dependence of the Coulomb peak splitting on top
gate voltage.} \label{fig2}
\end{figure}

The tunable double quantum dots are investigated in transport
measurements in a $^3$He/$^4$He-dilution refrigerator at a base
temperature of 50~mK.  We use a standard lock-in technique to
measure the differential conductance with an AC-excitation voltage
of 10~$\mu$V at a frequency of 16~Hz.

Figure \ref{fig2}(a) shows an image of the differential
conductance $G$ as a function of in-plane gate voltages $V_{G1}$
and $V_{G2}$. Both dots are weakly coupled to the reservoirs.  In
the low-conductance regions the electron number on dot $A$ is $m$
and on dot $B$ is $n$. $(m,n)$ is stable. The nearly horizontal
lines denote changes of the charge on dot $A$ by one electron,
i.~e.\ transitions from $(m,n)$ to $(m+1,n)$.  The nearly vertical
lines denote the respective transitions for dot $B$.  At the
intersections of both types of lines, both dots are in resonance
with each other.  This situation corresponds to transitions from
$(m+1,n)$ to $(m,n+1)$. In Figure \ref{fig2}(a), each of these
intersections is just split into two triple points indicating a
weak interdot coupling \cite{Wiel-03}. We have increased the top
gate voltage from $V_T=-105$~mV to $-103$~mV in Figure
\ref{fig2}(b). The splitting of the triple points has clearly
increased leading to more obvious hexagonal Coulomb blockade
domains. This indicates a stronger interdot coupling achieved by
tuning the top gate voltage $V_T$. The traces in Figures
\ref{fig2}(c) and \ref{fig2}(d) illustrate the increase of the
splitting of the triple points with increasing top gate voltage
(dashed lines in Figs.\ \ref{fig2}(a) and \ref{fig2}(b)).

The quantum dots are totally isolated from each other when we
apply a top-gate voltage $V_T<-110$~mV.  We then get the signature
of two non-interacting parallel quantum dots.  To completely
deplete the two-dimensional electron gas in absence of an oxidized
barrier a voltage of $V_T<-190$~mV is needed.  So, there still is
a common drain contact when the dots are isolated at
$V_T=-110$~mV. At more positive top gate voltages, the interdot
coupling increases. At $V_T = 0$, both dots have merged and we
observe one single large quantum dot. Thus, at intermediate
voltages the coupling of the two quantum dots is not purely
capacitively but at some point tunnel coupling sets on.

\begin{figure}
 \includegraphics{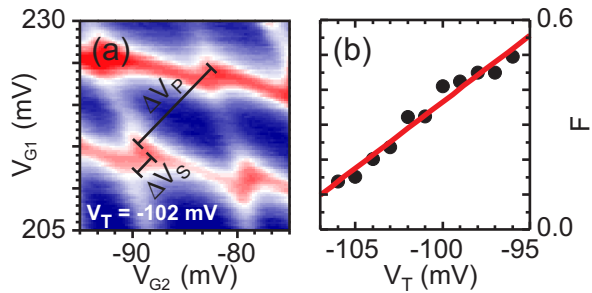}
 \caption{(a) Peak spacing $\Delta V_P$ and peak splitting $\Delta
V_S$ are determined from a grey scale plot of the differential
conductance $G$ as a function of gate voltages $V_{G2}$ and
$V_{G1}$. (b) Fractional peak splitting $F = 2 \Delta V_S / \Delta
V_P$ as a function of top gate voltage $V_T$.} \label{fig3}
\end{figure}

The dependence of the interdot coupling on top gate voltage is
further investigated in Figure \ref{fig3}. The coupling is
characterized by the fractional peak splitting $F = 2 \Delta V_S /
\Delta V_P$ with $\Delta V_S$ the Coulomb peak splitting and
$\Delta V_P$ the period \cite{Golden-96}.  For totally decoupled
dots $F$ is $0$ and for completely merged dots $F=1$.  We observe
a linear dependence of $F$ on $V_T$ ranging from $F=0.1$ at
$V_T=-106$~mV to $F=0.55$ at $V_T=-96$~mV in Figure \ref{fig3}(b).
This large range of fractional peak splitting observed nicely
demonstrates the tunability of the interdot coupling in our
double-dot system.

In conclusion, we have demonstrated the combination of local
anodic oxidation with an AFM and electron-beam lithography on a
GaAs/AlGaAs heterostructure.  We have shown that this combination
of well-known fabrication schemes allows an easy fabrication of
advantageous gate layouts.  We have fabricated two parallel
quantum dots with tunable interdot coupling.  The electronic
properties of this device were investigated in transport
measurements.  We have demonstrated the tunability of the interdot
coupling.

We thank F.~Hohls and U.~Zeitler for helpful discussions and help
with the measurement setup.  This work has been supported by BMBF.


\newpage


\begin{thebibliography}{20}
\expandafter\ifx\csname natexlab\endcsname\relax\def\natexlab#1{#1}\fi
\expandafter\ifx\csname bibnamefont\endcsname\relax
  \def\bibnamefont#1{#1}\fi
\expandafter\ifx\csname bibfnamefont\endcsname\relax
  \def\bibfnamefont#1{#1}\fi
\expandafter\ifx\csname citenamefont\endcsname\relax
  \def\citenamefont#1{#1}\fi
\expandafter\ifx\csname url\endcsname\relax
  \def\url#1{\texttt{#1}}\fi
\expandafter\ifx\csname urlprefix\endcsname\relax\def\urlprefix{URL }\fi
\providecommand{\bibinfo}[2]{#2}
\providecommand{\eprint}[2][]{\url{#2}}

\bibitem[{\citenamefont{Kouwenhoven et~al.}(1997)\citenamefont{Kouwenhoven,
  Marcus, McEuen, Tarucha, Westerveld, and Wingreen}}]{Kouwenhoven-97}
\bibinfo{author}{\bibfnamefont{L.~P.} \bibnamefont{Kouwenhoven}},
  \bibinfo{author}{\bibfnamefont{C.~M.} \bibnamefont{Marcus}},
  \bibinfo{author}{\bibfnamefont{P.~L.} \bibnamefont{McEuen}},
  \bibinfo{author}{\bibfnamefont{S.}~\bibnamefont{Tarucha}},
  \bibinfo{author}{\bibfnamefont{R.~M.} \bibnamefont{Westerveld}},
  \bibnamefont{and} \bibinfo{author}{\bibfnamefont{N.~S.}
  \bibnamefont{Wingreen}}, in \emph{\bibinfo{booktitle}{Mesoscopic Electron
  Transport}}, edited by \bibinfo{editor}{\bibfnamefont{L.~L.}
  \bibnamefont{Sohn}}, \bibinfo{editor}{\bibfnamefont{L.~P.}
  \bibnamefont{Kouwenhoven}}, \bibnamefont{and}
  \bibinfo{editor}{\bibfnamefont{G.}~\bibnamefont{Sch\"o{}n}}
  (\bibinfo{publisher}{Kluwer}, \bibinfo{address}{Dordrecht},
  \bibinfo{year}{1997}), vol. \bibinfo{volume}{345} of
  \emph{\bibinfo{series}{Series E}}, pp. \bibinfo{pages}{105--214}.

\bibitem[{\citenamefont{Kastner}(1992)}]{Kastner-92}
\bibinfo{author}{\bibfnamefont{M.~A.} \bibnamefont{Kastner}},
  \bibinfo{journal}{Rev. Mod. Phys.} \textbf{\bibinfo{volume}{64}},
  \bibinfo{pages}{849} (\bibinfo{year}{1992}).

\bibitem[{\citenamefont{van~der Wiel et~al.}(2003)\citenamefont{van~der Wiel,
  Franceschi, Elzerman, Fujisawa, Tarucha, and Kouwenhoven}}]{Wiel-03}
\bibinfo{author}{\bibfnamefont{W.~G.} \bibnamefont{van~der Wiel}},
  \bibinfo{author}{\bibfnamefont{S.~D.} \bibnamefont{Franceschi}},
  \bibinfo{author}{\bibfnamefont{J.~M.} \bibnamefont{Elzerman}},
  \bibinfo{author}{\bibfnamefont{T.}~\bibnamefont{Fujisawa}},
  \bibinfo{author}{\bibfnamefont{S.}~\bibnamefont{Tarucha}}, \bibnamefont{and}
  \bibinfo{author}{\bibfnamefont{L.~P.} \bibnamefont{Kouwenhoven}},
  \bibinfo{journal}{Rev. Mod. Phys.} \textbf{\bibinfo{volume}{75}},
  \bibinfo{pages}{1} (\bibinfo{year}{2003}).

\bibitem[{\citenamefont{Kouwenhoven et~al.}(2001)\citenamefont{Kouwenhoven,
  Austing, and Tarucha}}]{Kouwenhoven-01a}
\bibinfo{author}{\bibfnamefont{L.~P.} \bibnamefont{Kouwenhoven}},
  \bibinfo{author}{\bibfnamefont{D.~G.} \bibnamefont{Austing}},
  \bibnamefont{and} \bibinfo{author}{\bibfnamefont{S.}~\bibnamefont{Tarucha}},
  \bibinfo{journal}{Rep. Prog. Phys.} \textbf{\bibinfo{volume}{64}},
  \bibinfo{pages}{701} (\bibinfo{year}{2001}).

\bibitem[{\citenamefont{Blick et~al.}(1998)\citenamefont{Blick, Pfannkuche,
  Haug, v.~Klitzing, and Eberl}}]{Blick-98a}
\bibinfo{author}{\bibfnamefont{R.~H.} \bibnamefont{Blick}},
  \bibinfo{author}{\bibfnamefont{D.}~\bibnamefont{Pfannkuche}},
  \bibinfo{author}{\bibfnamefont{R.~J.} \bibnamefont{Haug}},
  \bibinfo{author}{\bibfnamefont{K.}~\bibnamefont{v.~Klitzing}},
  \bibnamefont{and} \bibinfo{author}{\bibfnamefont{K.}~\bibnamefont{Eberl}},
  \bibinfo{journal}{Phys. Rev. Lett.} \textbf{\bibinfo{volume}{80}},
  \bibinfo{pages}{4032} (\bibinfo{year}{1998}).

\bibitem[{\citenamefont{Holleitner et~al.}(2002)\citenamefont{Holleitner,
  Blick, H\"u{}ttel, Eberl, and Kotthaus}}]{Holleitner-02}
\bibinfo{author}{\bibfnamefont{A.~W.} \bibnamefont{Holleitner}},
  \bibinfo{author}{\bibfnamefont{R.~H.} \bibnamefont{Blick}},
  \bibinfo{author}{\bibfnamefont{A.~K.} \bibnamefont{H\"u{}ttel}},
  \bibinfo{author}{\bibfnamefont{K.}~\bibnamefont{Eberl}}, \bibnamefont{and}
  \bibinfo{author}{\bibfnamefont{J.~P.} \bibnamefont{Kotthaus}},
  \bibinfo{journal}{Science} \textbf{\bibinfo{volume}{297}},
  \bibinfo{pages}{70} (\bibinfo{year}{2002}).

\bibitem[{\citenamefont{Loss and DiVincenzo}(1998)}]{Loss-98}
\bibinfo{author}{\bibfnamefont{D.}~\bibnamefont{Loss}} \bibnamefont{and}
  \bibinfo{author}{\bibfnamefont{D.~P.} \bibnamefont{DiVincenzo}},
  \bibinfo{journal}{Phys. Rev. A} \textbf{\bibinfo{volume}{57}},
  \bibinfo{pages}{120} (\bibinfo{year}{1998}).

\bibitem[{\citenamefont{Hu and Sarma}(2000)}]{Hu-00}
\bibinfo{author}{\bibfnamefont{X.}~\bibnamefont{Hu}} \bibnamefont{and}
  \bibinfo{author}{\bibfnamefont{S.~D.} \bibnamefont{Sarma}},
  \bibinfo{journal}{Phys. Rev. A} \textbf{\bibinfo{volume}{61}},
  \bibinfo{pages}{062301} (\bibinfo{year}{2000}).

\bibitem[{\citenamefont{Golovach and Loss}(2002)}]{Golovach-02}
\bibinfo{author}{\bibfnamefont{V.~N.} \bibnamefont{Golovach}} \bibnamefont{and}
  \bibinfo{author}{\bibfnamefont{D.}~\bibnamefont{Loss}},
  \bibinfo{journal}{Semicon. Sci. Technol.} \textbf{\bibinfo{volume}{17}},
  \bibinfo{pages}{355} (\bibinfo{year}{2002}).

\bibitem[{\citenamefont{Vandersypen et~al.}(2002)\citenamefont{Vandersypen,
  Hanson, van Beveren, Elzerman, Greidanus, Franceschi, and
  Kouwenhoven}}]{Vandersypen-02}
\bibinfo{author}{\bibfnamefont{L.~M.~K.} \bibnamefont{Vandersypen}},
  \bibinfo{author}{\bibfnamefont{R.}~\bibnamefont{Hanson}},
  \bibinfo{author}{\bibfnamefont{L.~H.~W.} \bibnamefont{van Beveren}},
  \bibinfo{author}{\bibfnamefont{J.~M.} \bibnamefont{Elzerman}},
  \bibinfo{author}{\bibfnamefont{J.~S.} \bibnamefont{Greidanus}},
  \bibinfo{author}{\bibfnamefont{S.~D.} \bibnamefont{Franceschi}},
  \bibnamefont{and} \bibinfo{author}{\bibfnamefont{L.~P.}
  \bibnamefont{Kouwenhoven}}, in \emph{\bibinfo{booktitle}{Quantum Computing
  and Quantum Bits in Mesoscopic Systems}} (\bibinfo{publisher}{Kluwer},
  \bibinfo{address}{Dordrecht}, \bibinfo{year}{2002}).

\bibitem[{\citenamefont{Hofmann et~al.}(1995)\citenamefont{Hofmann, Heinzel,
  Wharam, Kotthaus, B\"ohm, Klein, Tr\"ankle, and Weimann}}]{Hofmann-95}
\bibinfo{author}{\bibfnamefont{F.}~\bibnamefont{Hofmann}},
  \bibinfo{author}{\bibfnamefont{T.}~\bibnamefont{Heinzel}},
  \bibinfo{author}{\bibfnamefont{D.~A.} \bibnamefont{Wharam}},
  \bibinfo{author}{\bibfnamefont{J.~P.} \bibnamefont{Kotthaus}},
  \bibinfo{author}{\bibfnamefont{G.}~\bibnamefont{B\"ohm}},
  \bibinfo{author}{\bibfnamefont{W.}~\bibnamefont{Klein}},
  \bibinfo{author}{\bibfnamefont{G.}~\bibnamefont{Tr\"ankle}},
  \bibnamefont{and} \bibinfo{author}{\bibfnamefont{G.}~\bibnamefont{Weimann}},
  \bibinfo{journal}{Phys. Rev. B} \textbf{\bibinfo{volume}{51}},
  \bibinfo{pages}{13872} (\bibinfo{year}{1995}).

\bibitem[{\citenamefont{Molenkamp et~al.}(1995)\citenamefont{Molenkamp,
  Flensberg, and Kemerink}}]{Molenkamp-95}
\bibinfo{author}{\bibfnamefont{L.~W.} \bibnamefont{Molenkamp}},
  \bibinfo{author}{\bibfnamefont{K.}~\bibnamefont{Flensberg}},
  \bibnamefont{and} \bibinfo{author}{\bibfnamefont{M.}~\bibnamefont{Kemerink}},
  \bibinfo{journal}{Phys. Rev. Lett.} \textbf{\bibinfo{volume}{75}},
  \bibinfo{pages}{4282} (\bibinfo{year}{1995}).

\bibitem[{\citenamefont{Adourian et~al.}(1999)\citenamefont{Adourian,
  Livermore, and Westervelt}}]{Adourian-99}
\bibinfo{author}{\bibfnamefont{A.~S.} \bibnamefont{Adourian}},
  \bibinfo{author}{\bibfnamefont{C.}~\bibnamefont{Livermore}},
  \bibnamefont{and} \bibinfo{author}{\bibfnamefont{R.~M.}
  \bibnamefont{Westervelt}}, \bibinfo{journal}{Appl. Phys. Lett.}
  \textbf{\bibinfo{volume}{75}}, \bibinfo{pages}{424} (\bibinfo{year}{1999}).

\bibitem[{\citenamefont{Holleitner et~al.}(2001)\citenamefont{Holleitner,
  Decker, Qin, Eberl, and Blick}}]{Holleitner-01}
\bibinfo{author}{\bibfnamefont{A.~W.} \bibnamefont{Holleitner}},
  \bibinfo{author}{\bibfnamefont{C.~R.} \bibnamefont{Decker}},
  \bibinfo{author}{\bibfnamefont{H.}~\bibnamefont{Qin}},
  \bibinfo{author}{\bibfnamefont{K.}~\bibnamefont{Eberl}}, \bibnamefont{and}
  \bibinfo{author}{\bibfnamefont{R.~H.} \bibnamefont{Blick}},
  \bibinfo{journal}{Phys. Rev. Lett.} \textbf{\bibinfo{volume}{87}},
  \bibinfo{pages}{256802} (\bibinfo{year}{2001}).

\bibitem[{\citenamefont{Fujita et~al.}(1996)\citenamefont{Fujita, Ohnishi,
  Ochiai, and Matsui}}]{Fujita-96}
\bibinfo{author}{\bibfnamefont{J.}~\bibnamefont{Fujita}},
  \bibinfo{author}{\bibfnamefont{Y.}~\bibnamefont{Ohnishi}},
  \bibinfo{author}{\bibfnamefont{Y.}~\bibnamefont{Ochiai}}, \bibnamefont{and}
  \bibinfo{author}{\bibfnamefont{S.}~\bibnamefont{Matsui}},
  \bibinfo{journal}{Appl. Phys. Lett.} \textbf{\bibinfo{volume}{68}},
  \bibinfo{pages}{1297} (\bibinfo{year}{1996}).

\bibitem[{\citenamefont{Ishii and Matsumoto}(1995)}]{Ishii-95}
\bibinfo{author}{\bibfnamefont{M.}~\bibnamefont{Ishii}} \bibnamefont{and}
  \bibinfo{author}{\bibfnamefont{K.}~\bibnamefont{Matsumoto}},
  \bibinfo{journal}{Jpn. J. Appl. Phys.} \textbf{\bibinfo{volume}{34}},
  \bibinfo{pages}{1329} (\bibinfo{year}{1995}).

\bibitem[{\citenamefont{Held et~al.}(1998)\citenamefont{Held, Vancura, Heinzel,
  Ensslin, Molland, and Wegscheider}}]{Held-98}
\bibinfo{author}{\bibfnamefont{R.}~\bibnamefont{Held}},
  \bibinfo{author}{\bibfnamefont{T.}~\bibnamefont{Vancura}},
  \bibinfo{author}{\bibfnamefont{T.}~\bibnamefont{Heinzel}},
  \bibinfo{author}{\bibfnamefont{K.}~\bibnamefont{Ensslin}},
  \bibinfo{author}{\bibfnamefont{M.}~\bibnamefont{Molland}}, \bibnamefont{and}
  \bibinfo{author}{\bibfnamefont{W.}~\bibnamefont{Wegscheider}},
  \bibinfo{journal}{Appl. Phys. Lett.} \textbf{\bibinfo{volume}{73}},
  \bibinfo{pages}{262} (\bibinfo{year}{1998}).

\bibitem[{\citenamefont{Keyser et~al.}(2000)\citenamefont{Keyser, Schumacher,
  Zeitler, Haug, and Eberl}}]{Keyser-00}
\bibinfo{author}{\bibfnamefont{U.~F.} \bibnamefont{Keyser}},
  \bibinfo{author}{\bibfnamefont{H.~W.} \bibnamefont{Schumacher}},
  \bibinfo{author}{\bibfnamefont{U.}~\bibnamefont{Zeitler}},
  \bibinfo{author}{\bibfnamefont{R.~J.} \bibnamefont{Haug}}, \bibnamefont{and}
  \bibinfo{author}{\bibfnamefont{K.}~\bibnamefont{Eberl}},
  \bibinfo{journal}{Appl. Phys. Lett.} \textbf{\bibinfo{volume}{76}},
  \bibinfo{pages}{457} (\bibinfo{year}{2000}).

\bibitem[{\citenamefont{Fuhrer et~al.}(2001)\citenamefont{Fuhrer, L\"u{}scher,
  Ihn, Heinzel, Ensslin, Wegscheider, and Bichler}}]{Fuhrer-01b}
\bibinfo{author}{\bibfnamefont{A.}~\bibnamefont{Fuhrer}},
  \bibinfo{author}{\bibfnamefont{S.}~\bibnamefont{L\"u{}scher}},
  \bibinfo{author}{\bibfnamefont{T.}~\bibnamefont{Ihn}},
  \bibinfo{author}{\bibfnamefont{T.}~\bibnamefont{Heinzel}},
  \bibinfo{author}{\bibfnamefont{K.}~\bibnamefont{Ensslin}},
  \bibinfo{author}{\bibfnamefont{W.}~\bibnamefont{Wegscheider}},
  \bibnamefont{and} \bibinfo{author}{\bibfnamefont{M.}~\bibnamefont{Bichler}},
  \bibinfo{journal}{Phys. Rev. B} \textbf{\bibinfo{volume}{63}},
  \bibinfo{pages}{125309} (\bibinfo{year}{2001}).

\bibitem[{\citenamefont{Golden and Halperin}(1996)}]{Golden-96}
\bibinfo{author}{\bibfnamefont{J.~M.} \bibnamefont{Golden}} \bibnamefont{and}
  \bibinfo{author}{\bibfnamefont{B.~I.} \bibnamefont{Halperin}},
  \bibinfo{journal}{Phys. Rev. B} \textbf{\bibinfo{volume}{54}},
  \bibinfo{pages}{16757} (\bibinfo{year}{1996}).

\end{thebibliography}



\end{document}